\documentclass[preprint,aps,12pt,showpacs,nofootinbib,tightenlines]{revtex4}
\usepackage{mathrsfs}
\usepackage{amsmath}
\usepackage{amssymb}
\usepackage{epsfig}
\usepackage{graphicx}
\usepackage{multirow}
\newcommand{\psl}{ p \hspace{-2.4truemm}/ }

\def\be{\begin{eqnarray}}
\def\en{\end{eqnarray}}
\def\non{\nonumber\\}

\def\prd{{Phys. Rev. D}~}
\def\prl{{ Phys. Rev. Lett.}~}
\def\plb{{ Phys. Lett. B}~}

\begin{document}
\title{$B_{(s)}\to D^*_{s0}(2317)P(V)$ decays in perturbative QCD approach}
\author{Zhi-Qing Zhang$^1$, Hongxia Guo\footnote{Corresponding author: Hongxia Guo, e-mail: guohongxia@zzu.edu.cn.}$^2$, Na Wang$^1$,
Hai-Tao Jia$^1$}
\affiliation{\it \small $^1$  Department of Physics, Henan
University of
Technology,\\ \small \it Zhengzhou, Henan 450052, P. R. China; \\
\small $^2$ \it School of Mathematics and Statistics, Zhengzhou University,\\
\small \it Zhengzhou, Henan 450001, P. R. China } 
\date{\today}
\begin{abstract}
In this work, we use pQCD approach to calculate 20 $B_{(s)}\to D^*_{s0}(2317)P(V)$
two body decays by assuming $D^*_{s0}(2317)$ as a $\bar cs$ scalar meson, where $P(V)$ denotes a pseudoscalar (vector) meson. These $B_{(s)}$
decays can serve as an ideal platform to probe the valuable information on the inner
structure of the charmed-strange meson $D^*_{s0}(2317)$, and to explore the dynamics of strong
interactions and signals of new physics. These considered decays can be divided into two
types: the CKM favored decays and the CKM suppressed decays.
The former are induced by $b\to c$ transition, whose branching ratios are larger than $10^{-5}$.
The branching fraction of the decay $\bar B^0_s\to D^{*+}_{s0}(2317)\rho^{-}$ is the largest
and reaches about $1.8\times 10^{-3}$, while the branching ratios for the decay
$\bar B^0_s\to D^{*+}_{s0}(2317)K^{*-}$ and other two pure annihilation decays $\bar B^0\to D^{*+}_{s0}(2317)K^-,
D^{*+}_{s0}(2317)K^{*-}$ are only at $10^{-5}$ order. Our predictions are consistent well with the results given by the light
cone sum rules approach. These decays are most likely to be  measured at the running LHCb and the forthcoming SuperKEKB.
The latter are induced by $b\to u$ transition, among of which
the channel $\bar B^0\to D^{*-}(2317)\rho^+$ has the largest branching fraction, reaching up to $10^{-5}$ order. Again the pure
annihilation decays $B^-\to D^{*-}_{s0}(2317)\phi, \bar B^0\to D^{*-}_{s0}(2317)K^+(K^{*+}),
B^-\to D^{*-}_{s0}(2317)K^0(K^{*0})$, have the smallest branching ratios, which drop to as low as $10^{-10}\sim10^{-8}$.
\end{abstract}

\pacs{13.25.Hw, 12.38.Bx, 14.40.Nd}
\vspace{1cm}

\maketitle

\section{Introduction}\label{intro}
The charmed-strange meson $D^*_{s0}(2317)$ was first observed by BABAR Collaboration in the inclusive
$D^+_s\pi^0$ invariant mass distribution \cite{babar1,babar2}, then confirmed by CLEO \cite{cleo}
and Belle Collaboration \cite{belle}, respectively. Usually, the $D^*_{s0}(2317)$ meson is suggested
as a P-wave $\bar cs$ state with spin-parity $J^P=0^+$. However, there exit two divergences between
the data and the theoretical predictions:
First, the measured mass for this meson is at least $150 MeV/c^2$ lower than
the theoretical calculations from a potential model \cite{god,kala}, lattice QCD \cite{bali} and so on.
For example, the authors \cite{haya} obtained $M(D^*_{s0}(2317))=(2480\pm30)MeV$ by using the standard Borel-transformed QCD sum rule which was higher than the BABAR result by about $160$ MeV.
While, Narsion \cite{narison} used the QCD spectral sum rules to get $M(D^*_{s0}(2317))=(2297\pm113)$ MeV and reached the conclusion that $D_s(2317)$ is a $\bar cs$ state. Second, the absolute branching ratio of decay
$D^*_{s0}(2317)^{\pm}\to D^{\pm}\pi^0$ measured by BESIII Collaboration \cite{bes3} showed that
$D^*_{s0}(2317)^-$ tends to have a significantly
larger branching ratio to $\pi^0D^{*-}_s$ than to $\gamma D^{*-}_s$, which differs from the expectation of the
conventional $\bar cs$ hypothesis. These puzzles inspired various exotic explanations to its inner structure,
such as $DK$ molecule state \cite{barn,chenyq,guo,faes,guo1}, a tetraquark state \cite{hycheng,tera,dmi,jrzhang}, or a mixture of a
$\bar c s$ state and a tetraquark state \cite{beve,mohler,liul}. In order to further reveal the internal structure
of $D^*_{s0}(2317)$, we intend to study the weak production of this charmed-strange meson through the $B_{(s)}$
decays, which can serve as an ideal platform to probe the valuable informations on the inner structure of
the exotic scalar mesons \cite{zqzhang1,zqzhang2,zqzhang3,zqzhang4,zou}. In the conventional two quark picture the branching ratios of the
decays $B_{(s)}\to D^*_{s0}(2317)P(V)$, where $P$ ($V$) denotes the light pseudoscalar
(vector) meson, are expected to be of the same order of magnitude as those of $B_{(s)}\to D_{s}P(V)$ decays, since the $D^*_{s0}(2317)$ meson
decay constant should be close to that of the pseudoscalar meson $D_s$ as required by the chiral symmetry. On the contrary, in the unconventional
picture the corresponding decay amplitudes involve additional hard scattering with the participation of four valence quarks. Then the branching ratios are at
least suppressed by the coupling constant and by inverse powers of heavy meson masses, such that they are much smaller than those of $B_{(s)}\to D_{s}P(V)$ decays
by one order. So it is meaningful to study the branching ratios of the decays $B_{(s)}\to D^*_{s0}(2317)P(V)$ both in experiment and theory.

$B_{(s)}$ two body nonleptonic decays with $D^*_{s0}(2317)$ meson involved in the final states have been studied
in the light cone sum rules (LCSR) approach \cite{lirh},
the relativistic quark model (RQM) \cite{faus}, and the nonrelativistic quark model (NRQM) \cite{albert}. Here we would like to use
pQCD approach to study $B_{(s)}\to D^*_{s0}(2317)P(V)$ decays. Studying these decays may shed light on the nature of the $D^*_{s0}(2317)$ meson,
explore the dynamics of strong interactions. Further more, the study of these  weak decays is important for further improvement
in the determination of the Cabibbo-Kobayshi-Maskawa (CKM) matrix elements, for testing the prediction of the Standard Model and searching for possible
deviations from theoretical predictions, the so-called "new physics" signals.

The layout of this paper is as follows, we analyze the decay $B_{(s)}\to D^*_{s0}(2317)P(V)$ using the perturbative QCD approach in Section II.
The numerical results and discussions are given in
Section III, where the theoretical uncertainties are also
considered. The conclusions are presented in the final part.
\section{the perturbative calculations}
In the pQCD approach, the only non-perturbative inputs are the light cone distribution amplitudes (LCDAs) and the meson
decay constants. For the wave function of the heavy $B_{(s)}$ meson,
we take
\be
\Phi_{B_{(s)}}(x,b)=
\frac{1}{\sqrt{2N_c}} (\psl_{B_{(s)}} +m_{B_{(s)}}) \gamma_5 \phi_{B_{(s)}} (x,b).
\label{bmeson}
\en
Here only the contribution of Lorentz structure $\phi_{B_{(s)}} (x,b)$ is taken into account, since the contribution
of the second Lorentz structure $\bar \phi_{B_{(s)}}$ is numerically small \cite{cdlu} and has been neglected. For the
distribution amplitude $\phi_{B_{(s)}}(x,b)$ in Eq.(\ref{bmeson}), we adopt the following model:
\be
\phi_{B_{(s)}}(x,b)=N_{B_{(s)}}x^2(1-x)^2\exp[-\frac{M^2_{B_{(s)}}x^2}{2\omega^2_b}-\frac{1}{2}(\omega_bb)^2],
\en
where $\omega_b$ is a free parameter, we take $\omega_b=0.4\pm0.04 (0.5\pm0.05)$ GeV for $B(B_s)$ meson in numerical calculations, and $N_B=101.445(N_{B_s}=63.671)$
is the normalization factor for $\omega_b=0.4(0.5)$. These parameters has been fixed using the rich experimental data on the
$B_{(s)}$ decay channels. In this model the significant feature is the intrinsic transverse momentum dependence, which is essential for the $B_{(s)}$ meson.
It can It can provide additional suppression in the large $b$ region, where the soft dynamics dominates and Sudakov
suppression is weaker. Considering a small SU(3) breaking, the $s$ quark momentum fraction is a litter larger than that of the $u(d)$ quark in the lighter $B$ meson, because of the heavier
mass for the $s$ quark. From the shape of the distribution amplitude shown in Ref.\cite{ali}, it is easy to see that the larger $\omega_b$ gives a larger momentum fraction
to the $s$ quark.

The wave functions of the scalar meson $D^*_{s0}$ \footnote{From now on, we will use $D^*_{s0}$ to denote $D^*_{s0}(2317)$ for simply
in some places.}, we use the form defined in Ref.\cite{chench}
\be
\langle\bar D^{*}_{s0}(2317)^+(p_2)|\bar c_\beta(z)s_{\gamma}(0)|0\rangle&=&\frac{1}{\sqrt{2N_c}}\int dxe^{ip_2\cdot z}\left[(\psl_{2})_{lj}+m_{D^*_{s0}}I_{lj}\right]\phi_{D^*_{s0}}.
\en
It is noticed that the distribution amplitudes which associate with the nonlocal operators $\bar c(z)\gamma_\mu s$
and $\bar c(z)s$ are different. The difference between them is order of $\bar \Lambda/m_{D^*_{s0}}\sim(m_{D^*_{s0}}-m_c)/m_{D^*_{s0}}$.
If we set $m_{D^*_{s0}}\sim m_c$, we can get these two distribution amplitudes are very similar. For
the leading power calculation, it is reasonable to parameterize them in the same form as
\be
\phi_{D^*_{s0}}(x)=\tilde{f}_{D^*_{s0}}6x(1-x)\left[1+a(1-2x)\right]
\en
in the heavy quark limit. Here $\tilde{f}_{D^*_{s0}}=225\pm25$ MeV is determined from the two-point
QCD sum rules, and the shape parameter $a=-0.21$ \cite{lirh} is fixed under the condition that the distribution
amplitude $\phi_{D^*_{s0}}(x)$ possesses the maximum at $\bar x=(m_{D^*_{s0}}-m_c)/m_{D^*_{s0}}$ with
$m_c=1.275$ GeV. It is worthwhile to point out that the intrinsic $b$ dependence of this charmed meson's
wave functions has been neglected in our analysis.

Since the light cone distribution amplitudes  of the pseudoscalar mesons $\pi, K, \eta^{(\prime)}$ and the vector mesons $\rho, K^*, \omega$
have been well constrained in the papers \cite{wfun1,wfun2,wfun3,pball,pball1,pball2,hnli}, and been tested
systematically in the work \cite{ali}, we will use these LCDAs directly listed in that paper \cite{ali},
together with the corresponding decay constants.

For these processes considered, the weak effective Hamiltonian
$H_{eff}$ can be written as two types: \be
H_{eff}=\frac{G_F}{\sqrt{2}}V_{cb}V^*_{uq}[C_1(\mu)O_1(\mu)+C_2(\mu)O_2(\mu)],
\;\;\;\;\text{Type I} \en where the tree operators are given as:
\be O_1=(\bar c_{\alpha}b_{\beta})_{V-A}(\bar q_{\beta}u_{\alpha})_{V-A},\;\;\;
O_2=(\bar c_{\alpha}b_{\alpha})_{V-A}(\bar q_{\beta}u_{\alpha})_{V-A},
\en
for the CKM favored channels, while the effective Hamiltonian for the CKM suppressed decays is written as:
 \be
H_{eff}=\frac{G_F}{\sqrt{2}}V_{ub}V^*_{cq}[C_1(\mu)O_1(\mu)+C_2(\mu)O_2(\mu)],
\;\;\;\;\text{Type II} \en with \be O_1=(\bar
u_{\alpha}b_{\beta})_{V-A}(\bar D_{\beta}c_{\alpha})_{V-A},
O_2=(\bar u_{\alpha}b_{\alpha})_{V-A}(\bar
D_{\beta}c_{\alpha})_{V-A}. \en
Here $D$ represents $d(s)$ quark.  The type I channel is induced by $b\to c$
transition, such as $\bar B^0_s\to D^{*+}_{s0}\pi^{-}(K^-),D^{*+}_{s0}\rho^{-}(K^{*-}), \bar B^0\to D^{*+}_{s0}K^{-}(K^{*-})$.
While the type II decay is induced by $b\to u$ transition, such as $\bar B^0\to D^{*-}_{s0}\pi^+(K^+),
D^{*-}_{s0}\rho^{+}(K^{*+}), B^-\to D^{*-}_{s0}\pi^0(\rho^0,\omega,\phi), D^{*-}_{s0}\eta^{(\prime)}$.
\begin{figure}[t]
\vspace{-4cm} \centerline{\epsfxsize=18 cm \epsffile{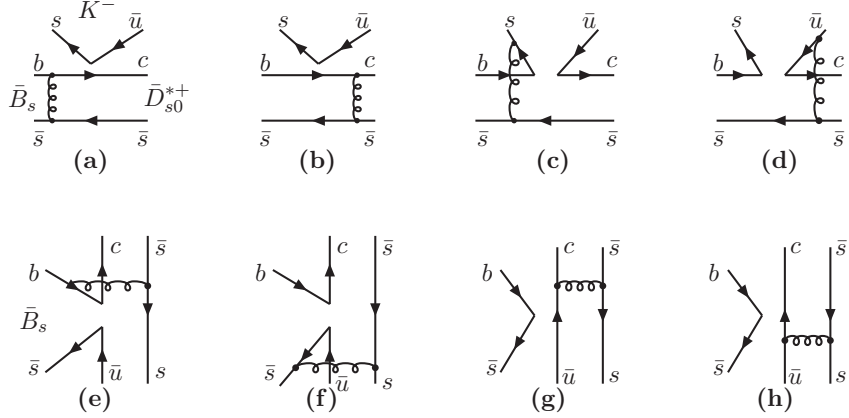}}
\vspace{-15.5cm} \caption{ Diagrams contributing to the $\bar B^0_s\to
D^{*+}_{s0}K^-$ decay.}
 \label{Figure1}
\end{figure}
For the CKM favored decays, we take the decay $\bar B^0_s\to D^{*+}_{s0}K^-$ as an example, whose Feynman diagrams
are given in Fig.1.
The first line Feynman diagrams are for the emission type ones, where Fig.1(a) and 1(b) are
the factorization diagrams, Fig.1(c)
and 1(d) are the nonfactorization ones, their amplitudes can be written as:
\be
\mathcal{F}^{P}_{ B\to D^*_{s0}}&=&8\pi C_FM^4_{B}f_P\int_0^1 d x_{1} dx_{2}\,
\int_{0}^{\infty} b_1 db_1 b_2 db_2\, \phi_{B}(x_1,b_1)\phi_{D^*_{s0}}\non && \times
\left\{[1+x_2-r_{D^*_{s0}}(2x_2-1)]
E_e(t_a)S_t(x_2)(t_a)h_e(x_1,x_2(1-r^2_{D^*_{s0}}),b_1,b_2)\right.\non &&\left.\left.
+[(2-r_{D^*_{s0}})r_{D^*_{s0}}+r_c(1-2r_{D^*_{s0}})]E_e(t_b)S_t(x_1)
h_e(x_2,x_1(1-r^2_{D^*_{s0}}),b_2,b_1)\right]\right\},\;\;\;\; \label{fe1}\\
\mathcal{M}^{P}_{ B\to D^*_{s0}}&=&-32\pi C_f m_{B}^4/\sqrt{2N_C}\int_0^1 d x_{1} dx_{2} dx_{3}
\int_{0}^{\infty} b_1 db_1 b_3
db_3\,\phi_{B}(x_1,b_1)\non &&
\times\phi_{D^*_{s0}}(x_2)\phi_{P}(x_3)\left[\left(r_{D^*_{s0}}x_2-x_3\right)E_{en}(t_c)h^{c}_{en}(x_1,x_2,x_3,b_1,b_3) \right.\non &&\left.
 +\left(1+x_2-x_3-r_{D^*_{s0}}x_2\right)E_{en}(t_{d})h^{d}_{en}(x_1,x_2,x_3,b_1,b_3)\right], \label{nfe1}
\en
where $P$ denotes a pseudoscalar meson, $r_{D^*_{s0}}=m_{D^*_{s0}}/M_{B}, r_{c}=m_c/M_{B}$ and $f_P$ is the decay constant
of the pseudoscalar meson. The evolution factors  evolving the scale $t$ and the hard functions for the hard part
of the amplitudes are listed as:
\be
E_e(t)&=&\alpha_s(t)\exp[-S_{B}(t)-S_{D^*_{s0}}(t)],\label{suda1}\\
E_{en}(t)&=&\alpha_s(t)\exp[-S_{B}(t)-S_{P}(t)-S_{D^*_{s0}}(t)|_{b_1=b_2}],\label{suda2}\\
h_e(x_1,x_2,b_1,b_2)&=&K_0(\sqrt{x_1x_2}m_{B}b_1)\left[\theta(b_1-b_2)K_0(\sqrt{x_2}m_{B}b_1)
I_0(\sqrt{x_2}m_{B}b_2)\right.
\non && \left.+\theta(b_2-b_1)K_0(\sqrt{x_2}m_{B}b_2)I_0(\sqrt{x_2}m_{B}b_1)\right],\\
h^{j}_{en}(x_1,x_2,x_3,b_1,b_3)&=&\left[\theta(b_1-b_3)K_0(\sqrt{A^2}m_{B}b_1)
I_0(\sqrt{A^2}m_{B}b_3)\right.\non && \left.+(b_1\leftrightarrow b_3)\right]
\left(\begin{matrix}K_0(A_jm_{B}b_3)& \text{for} A^2_j\geq 0\\
\frac{i\pi}{2}H^{(1)}_0(\sqrt{|A^2_j|}m_{B}b_3)&\text{for} A^2_j\leq 0\\\end{matrix}\right)|_{j=c,d},
\en
with the variables
\be
A^2&=&x_2x_1, \\
A^2_c&=&x_2(x_1-x_3(1-r^{2}_{D^*_{s0}})), \label{ac1}\\
A^2_{d}&=&x_2(x_1-(1-x_3)(1-r^2_{D^*_{s0}})).\label{ad1}
\en
The hard scales $t$  and the expression of Sudakov factor in each amplitude can be found in Appendix.
As we know that the double logarithms $\alpha_sln^2x$ produced by the radiative corrections are not
small expansion parameters when the end point region is important, in order to improve the
perturbative expansion, the threshold resummation of these logarithms to all order is needed, which
leads to a quark jet function
\be
S_t(x)=\frac{2^{1+2c}\Gamma(3/2+c)}{\sqrt{\pi}\Gamma(1+c)}[x(1-x)]^c,
\en
with $c=0.5$. It is effective to smear the end point singularity with a momentum fraction $x\to0$.
This factor will also appear in the factorizable annihilation amplitudes.

As to the amplitudes for the second line Feynman diagrams can be obtained by the Feynman rules
and are given as:
\be
\mathcal{M}^{D^*_{s0}}_{ann}&=&32\pi C_f m_{B}^4/\sqrt{2N_C}\int_0^1 d x_{1} dx_{2} dx_{3}\,\int_{0}^{\infty} b_1 db_1 b_3 db_3\,
\phi_{B}(x_1,b_1)\phi_{D^*_{s0}}(x_2)\non &&
\times\left\{-\left[r_{D^*_{s0}}r_P((x_2-x_3+3)\phi_P^p(x_3)
-(x_2+x_3-1)\phi^T_P(x_3))\right.\right.\non &&\left.\left.
+x_2\phi_P(x_3)\right]
E_{an}(t_{e})h^e_{an}(x_1,x_2,x_3,b_1,b_3)\right.\non &&\left.+\left[(1-x_3)\phi_P(x_3)
+r_{D^*_{s0}}r_P((x_2-x_3+1)\phi_P^p(x_3)+(x_2+x_3-1)\phi_P^T(x_3))\right]\right.\non &&\left.
\times E_{an}(t_{f})h^f_{an}(x_1,x_2,x_3,b_1,b_3)\right\}, \label{fe3}
\en
\be
\mathcal{F}^{D^*_{s0}}_{ann}&=&-8\pi C_f f_{B}m_{B}^4\int_0^1 d x_{2} dx_{3}\,
\int_{0}^{\infty} b_2 db_2 b_3 db_3\, \phi_{D^*_{S0}}(x_2)\left\{\left[4r_Pr_{D^*_{s0}}(1-x_3)\phi^p_P(x_3)\right.\right.\non &&
\left.\left.+r_P(r_c+2r_{D^*_{S0}}x_3)(\phi^p_P(x_3)+\phi^T_P(x_3))+(1+2r_cr_{D^*_{s0}}-x_3)\phi_P(x_3)\right]E_{af}(t_g) \right.\non &&\left.
\times S_t(x_3)h_{af}(x_2,x_3(1-r_{D^*_{s0}}^2),b_2,b_3)
-\left[x_2\phi_P(x_3)+2r_Pr_{D^*_{s0}}(1+x_2)\phi^p_P(x_3)\right]\right.\non &&\left.
\times E_{af}(t_h)S_t(x_2)h_{af}(x_3,x_2(1-r_{D^*_{s0}}^2),b_3,b_2)\right\}. \label{fe4}
\en
Here $\mathcal{F}^{D^*_{s0}}_{ann}(\mathcal{M}^{D^*_{s0}}_{ann})$ are the (non)factorizable annihilation type amplitudes, where the
evolution factors $E$ evolving the scale $t$ and the hard functions of the hard part
of factorization amplitudes are listed as:
\be
E_{an}(t)&=&\alpha_s(t)\exp[-S_{B}(t)-S_{D^*_{s0}}(t)-S_{P}(t)|_{b_2=b_3}],\label{suda3}\\
E_{af}(t)&=&\alpha_s(t)\exp[-S_{D^*_{s0}}(t)-S_{P}(t)],\label{suda4}\\
h_{an}^j(x_{i=1,2,3},b_1,b_3)&=&i\frac{\pi}{2}\left[\theta(b_1-b_3)H^{(1)}_0(\sqrt{x_2x_3(1-r^2_{D^*_{s0}})}m_{B}b_1)
J_0(\sqrt{x_2x_3(1-r^2_{D^*_{s0}})}m_{B}b_3)\right.\non &&\left.+(b_1\leftrightarrow b_3)\right]
\left(\begin{matrix}K_0(L_jm_{B}b_1)& \text{for} L^2_j\geq 0\\
\frac{i\pi}{2}H^{(1)}_0(\sqrt{|L^2_j|}m_{B}b_1)& \text{for} L^2_j\leq 0\\ \end{matrix}\right)|_{j=e,f},\\
h_{af}(x_2,x_3,b_2,b_3)&=&(i\frac{\pi}{2})^2H^{(1)}_0(\sqrt{x_2x_3}m_{B}b_2)\non &&
\times[\theta(b_2-b_3)H^{(1)}_0(\sqrt{x_3}m_{B}b_2)J_0(\sqrt{x_3}m_{B}b_3)+(b_2\leftrightarrow b_3)],
\en
where the definition of $L^2_j$ are written as:
\be
L^2_e&=&r_b^2-(1-x_2)\left[x_3(1-r^2_{D^*_{s0}})+r^2_{D^*_{s0}}-x_1\right],\label{lef1}\\
L^2_f&=&x_2\left[x_1-(1-x_3)(1-r^2_{D^*_{s0}})\right].\label{lef2} \en The functions
$H^{(1)}_0,J_0, K_0, I_0$ in the upper hard kernels
$h_{e},h^j_{en},h^j_{an},h_{af}$ are the (modified) Bessel
functions, which can be obtained from the Fourier transformations of
the quark and gluon propagators.

Similarly, we can also give the amplitudes for the CKM suppressed decay
channels,
\be
\mathcal{F}^{D^*_{s0}}_{B\to P}&=&8\pi C_FM^4_{B}f_{D^*_{S0}}\int_0^1 d x_{1} dx_{3}\,
\int_{0}^{\infty} b_1 db_1 b_3 db_3\, \phi_B(x_1,b_1)\non && \times
[(1+x_3)\phi_P(x_3)-r_P(2x_3-1)(\phi^p_P(x_3)+\phi^T_P(x_3))]\non &&
\times E_e(t_a)S_t(x_3)(t_a)h_e(x_1,x_3(1-r^2_{D}),b_1,b_3)\non &&
+[2r_P\phi_P^p(x_3)]E_e(t_b)S_t(x_1)
h_e(x_3,x_1(1-r^2_{D}),b_3,b_1)], \label{ckms1}\\
\mathcal{M}^{D^*_{s0}}_{B\to P}&=&32\pi C_f m_B^4/\sqrt{2N_C}\int_0^1 d x_{1} dx_{2} dx_{3}\,
\int_{0}^{\infty} b_1 db_1 b_2
db_2\,\phi_B(x_1,b_1)\phi_{D^*_{s0}}(x_2)\non &&
\times\left\{\left[x_2\phi_P(x_3)+r_Px_3(\phi^T_P(x_3)-\phi^p_P(x_3))\right]E_{en}(t_c)h^{c}_{en}(x_1,x_3,x_3,b_1,b_3) \right.\non &&\left.
 -\left[(1-x_2+x_3)\phi_P(x_3)-r_Px_3(\phi^T_P(x_3)+\phi^p_P(x_3))\right]\right.\non &&\left.\times E_{en}(t_{D^*_{s0}})h^{d}_{en}(x_1,x_2,x_3,b_1,b_3)\right\},
\en
where these two amplitudes are factorizable and nonfactorizable emission contributions, respectively. The amplitudes
$\mathcal{F}^{P}_{ B_s\to D^*_{s0}}, \mathcal{M}^{P}_{ B_s\to D^*_{s0}}$ are the color allowed amplitudes, while
$\mathcal{F}^{D^*_{s0}}_{B\to P}, \mathcal{M}^{D^*_{s0}}_{B\to P}$ are the color suppressed ones. The annihilation
type amplitudes are listed as:
\be
\mathcal{M}^{P}_{ann}&=&32\pi C_f m_B^4/\sqrt{2N_C}\int_0^1 d x_{1} dx_{2} dx_{3}\,\int_{0}^{\infty} b_1 db_1 b_2 db_2\,
\phi_B(x_1,b_1)\phi_{D^*_{S0}}(x_2)\non &&
\times\left\{\left[r_{D^*_{S0}}r_P((x_2+x_3+2)\phi_P^p(x_3)
-(x_2-x_3)\phi^T_P(x_3))\right.\right.\non &&\left.\left.
-x_3\phi_P(x_3)\right]
E_{an}(t_{e})h^e_{an}(x_1,x_2,x_3,b_1,b_3)\right.\non &&\left.+\left[-r_{D^*_{S0}}r_P((x_2+x_3)\phi_P^p(x_3)
+(x_2-x_3)\phi^T_P(x_3))+x_2\phi_P(x_3)\right]\right.\non &&\left. \times E_{an}(t_{f})h^f_{an}(x_1,x_2,x_3,b_1,b_3)\right\},
\en
\be
\mathcal{F}^{P}_{ann}&=&8\pi C_f f_{B}m_B^4\int_0^1 d x_{2} dx_{3}\,
\int_{0}^{\infty} b_2 db_2 b_3 db_3\, \phi_{D^*_{S0}}(x_2)\non &&\left\{\left[2r_{D^*_{S0}}r_P(1+x_2)\phi_P^p(x_3)-x_2\phi_P(x_3)\right]
E_{af}(t_g)S_t(x_3)h_{af}(x_3,x_2(1-r_{D^*_{S0}}^2),b_3,b_2)
\right.\non &&\left.+ \left[(x_3-r_{D^*_{S0}}(2r_c-r_{D^*_{S0}}))\phi_P(x_3)+r_P(r_c-2r_{D^*_{S0}}(1+x_3))\phi^p_P(x_3)\right.\right.
\non && \left.\left.
-r_P(r_c-2r_{D^*_{S0}}(1-x_3))\phi^T_P(x_3)\right]E_{af}(t_h)S_t(x_2)h_{af}(x_2,x_3(1-r_{D^*_{S0}}^2),b_2,b_3)\right\}.\label{ckms4}
\en
The definitions for the evolution factors, the hard functions and the jet function $S_t(x)$
in Eqs.(\ref{ckms1})$\sim$(\ref{ckms4}) can be found in  Eqs.(\ref{fe1}),(\ref{nfe1}) and Eqs.(\ref{fe3}),(\ref{fe4}) with the different parameters in the
hard function $h^{c,d}_{en}, h^{e,f}_{an}$, which are listed as:
\be
A\to A^{\prime2}&=&x_1x_3(1-r^2_{D^*_{s0}}), \\
A^{2}_c\to A^{\prime2}_c&=&(x_1-x_2)x_3(1-r_{D^*_{s0}}^2),\label{ac2}\\
A^{2}_{d}\to A^{\prime2}_{d}&=&r^2_c-r^2_{D^*_{s0}}+(x_1+x_2)r^2_{D^*_{s0}}-(1-x_2-x_1)x_3(1-r^2_{D^*_{S0}}),\label{ad2}\\
L^{2}_e\to L^{\prime2}_e&=&r_b^2-(1-x_2)\left[1-x_3(1-r^2_{D^*_{S0}})-x_2r^2_{D^*_{S0}}-x_1\right],\label{lefp1}\\
L^{2}_f\to L^{\prime2}_f&=&x_2\left[x_1-x_3(1-r^2_{D^*_{s0}})-x_2r^2_{D^*_{s0}}\right].\label{lefp2}
\en
For the decays $B_{(s)}\to D^*_{s0}V$, their amplitudes can be obtained from the ones of decays $B_{(s)}\to D^*_{s0}P$ with following substitutions:
\be
\phi_P\to\phi_V, \phi^p_P\to\phi^s_V, \phi^t_P\to\phi^t_V, r_P\to -r_V, f_P\to f_V.
\en
Combining these amplitudes, one can
ease to write out the total decay amplitude of each considered
channel:
\be
\mathcal{A}(\bar B^0_s\to D^{*+}_{s0}\pi^{-})&=&\frac{G_F}{\sqrt2}V_{cb}V^*_{ud}(\emph{F}^{\pi}_{B_s\to D^*_{s0}}a_1
+\emph{M}^{\pi}_{B_s\to D^*_{s0}}C_1), \label{dspi}\\
\mathcal{A}(\bar B^0_s\to D^{*+}_{s0}K^-)&=&\frac{G_F}{\sqrt2}V_{cb}V^*_{us}(\emph{F}^{K}_{B_s\to D^*_{s0}}a_1
+\emph{M}^{K}_{B_s\to D^*_{s0}}C_1+\emph{M}^{D^*_{s0}}_{ann}C_2+\emph{F}^{D^*_{s0}}_{ann}a_2),\\
\mathcal{A}(\bar B^0\to D^{*+}_{s0}K^{-})&=&\frac{G_F}{\sqrt2}V^*_{cb}V_{ud}(\emph{M}^{D^*_{s0}}_{ann}C_2+\emph{F}^{D^*_{s0}}_{ann}a_2),\\
\emph{A}(\bar B^0\to D^{*-}_{s0}\pi^+)&=&\frac{G_F}{\sqrt2}V_{ub}V^*_{cs}(\emph{F}^{D^{*}_{s0}}_{B\to \pi}a_1
+\emph{M}^{D^{*-}_{s0}}_{B\to \pi}C_1),\\
\mathcal{A}(B^-\to D^{*-}_{s0}\pi^0)&=&\frac{G_F}{\sqrt2}V_{ub}V^*_{cs}\frac{1}{\sqrt2}(\emph{F}^{D^{*}_{s0}}_{B\to \pi}a_1
+\emph{M}^{D^{*-}_{s0}}_{B\to \pi}C_1),\\
\mathcal{A}(\bar B^0\to D^{*-}_{s0}K^+)&=&\frac{G_F}{\sqrt2}V_{ub}V^*_{cd}(\emph{M}^{K}_{ann}C_2+\emph{F}^{K}_{ann}a_2),\\
\mathcal{A}(B^-\to D^{*-}_{s0}K^0)&=&\frac{G_F}{\sqrt2}V_{ub}V^*_{cd}(\emph{M}^{K}_{ann}C_1+\emph{F}^{K}_{ann}a_1),\\
\mathcal{A}(B^-\to D^{*-}_{s0}\eta_{n \bar n})&=&\frac{G_F}{\sqrt2}V_{ub}V^*_{cs}(\emph{F}^{D^{*}_{s0}}_{B\to \eta_{n\bar n}}a_1
+\emph{M}^{D^{*-}_{s0}}_{B\to \eta_{n\bar n}}C_1),\;\;\;\;\;\\
\mathcal{A}(B^-\to D^{*-}_{s0}\eta_{s\bar s})&=&\frac{G_F}{\sqrt2}V_{ub}V^*_{cs}
\left(\emph{M}^{\eta_{s\bar s}}_{ann}C_1+\emph{F}^{\eta_{s\bar s}}_{ann}a_1\right),\label{dsss}
\en
where $\eta_{n\bar n}=\frac{1}{\sqrt2}(u\bar u+d\bar d)$ and $\eta_{s\bar s}$. The physical states
$\eta$ and $\eta'$ can be related to these two flavor states $\eta_{n\bar n}$ and $\eta_{s\bar s}$ through
the following mixing mechanism:
\be \left(\begin{matrix}
|\eta\rangle\\
|\eta\rangle
\end{matrix}\right)=\left(\begin{matrix}\cos\phi & -\sin\phi\\
                           \sin\phi & \cos\phi\end{matrix}\right)
                           \left(\begin{matrix}|\eta_{n\bar n}\rangle \\ |\eta_{s\bar s}\rangle\end{matrix}\right),
\label{mixing} \en
with the mixing angle $\phi=39.3^\circ\pm1.0^\circ$ \cite{feldmann}.
The formulae for the $B_{(s)}\to D^{*}_{s0}V$ can be obtained through the following substitutions in Eqs.(\ref{dspi})-(\ref{dsss}),
\be
\pi^{\pm}\to \rho^{\pm},\;\;\; \pi^{0}\to \rho^{0},\omega,\;\;\; K\to K^*, \;\;\; \eta_{s\bar s}\to\phi.
\en
\section{the numerical results and discussions}
We use the following input parameters in the numerical calculations \cite{pdg16,lirh}:
\be
f_B&=&190 MeV, f_{B_s}=230 MeV, M_B=5.28 GeV, M_{B_s}=5.37 GeV, \\
\tau_B^\pm&=&1.638\times 10^{-12} s,\tau_{B^0}=1.519\times 10^{-12} s, \tau_{B_s}=1.512\times 10^{-12} s,\\
M_{W}&=&80.42 GeV, M_{D^*_{s0}}=2.3177 GeV, \tilde f_{D^*_{s0}}=(225\pm25) MeV. \en
For the CKM matrix elements, we adopt the Wolfenstein parametrization with values from Particle Data Group (PDG) \cite{pdg16} $A=0.811\pm0.026,
\lambda=0.22506\pm0.00050,
\bar\rho=0.124^{+0.019}_{-0.018}$ and $\bar\eta=0.356\pm0.011$ .

In the $B_{(s)}$-rest frame, the decay rates of $B_{(s)}\to D^{*}_{s0}(2317)P(V)$
can be written as
\be \mathcal{BR}(B_{(s)}\to D^{*}_{s0}(2317)P(V))=\frac{\tau_{B_{(s)}}}{16\pi M_B}(1-r^2_{D^{*}_{s0}}){\cal A}, \en
where $\cal A$ is the total decay amplitude of each considered decay, which has been listed in Eqs.(\ref{dspi})-(\ref{dsss}).
The branching ratios for the CKM favored (Type I) decays are given in Table \ref{tab1}, where one can find that our predictions
are consistent well with those calculated in the light cone sum rules approach within errors. While our predictions
are smaller than the results given by the relativistic quark model (RQM) \cite{faus} and the nonrelativistic quark model
(NRQM)\cite{albert}, respectively. Especially, for the pure annihilation decay $\bar B^0_s\to D^{*+}_{s0}K^{*-}$, whose
branching fraction reaches up to $10^{-3}$ predicted by NRQM approach, it seems too large to be acceptable.
\begin{table}
\caption{Branching ratios ($\times10^{-4}$) of the CKM favored (Type I) decays obtained in the pQCD
approach (This work), where the errors for these
entries correspond to the uncertainties in the $w_{b}=0.4\pm0.04(0.5\pm0.05)$ for $B(B_s)$ meson,
the hard scale $t$ varying from $0.75t$ to $1.25t$, and the CKM matrix elements.
 In Ref.\cite{lirh}, the branching ratios
are calculated in the factorization assumption (FA) with the form factors obtained in the light cone sum rules (LCSR). We also list the results given
by the relativistic quark model (RQM) \cite{faus} and the nonrelativistic quark model
(NRQM)\cite{albert}, respectively. }
\begin{center}
\begin{tabular}{c|c|c|c|c}
\hline\hline Modes & This work & LCSR \cite{lirh} &RQM\cite{faus}&  NRQM\cite{albert}  \\
\hline
$\bar B^0_s\to D^{*}_{s0}(2317)^+\pi^{-}$&$5.49^{+2.64+0.41+0.35}_{-1.68-0.27-0.35}$&$5.2^{+2.5}_{-2.1}$ &$9$&$10$\\
$\bar B^0_s\to D^{*}_{s0}(2317)^+K^-$ &$0.51^{+0.06+0.01+0.01}_{-0.04-0.01-0.01}$  &$0.4^{+0.2}_{-0.2}$&$0.7$&$0.9$\\
$\bar B^0_s\to D^{*}_{s0}(2317)^+\rho^-$ &$17.7^{+8.5+1.3+1.2}_{-5.3-0.8-1.1}$   &$13^{+6}_{-5}$&22&$27$\\
$\bar B^0_s\to D^{*}_{s0}(2317)^+K^{*-}$    &$1.01^{+0.44+0.06+0.05}_{-0.31-0.06-0.07}$     &$0.8^{+0.4}_{-0.3}$&1.2&$16$\\
$\bar B^0\to D^{*}_{s0}(2317)^+K^-$&$0.18^{+0.06+0.01+0.01}_{-0.04-0.01-0.01}$&--&--&--\\
$\bar B^0\to D^{*}_{s0}(2317)^+K^{*-}$&$0.25^{+0.07+0.02+0.01}_{-0.06-0.02-0.02}$&--&--&--\\
\hline\hline
\end{tabular}\label{tab1}
\end{center}
\end{table}

The Belle Collaboration has measured the product of the branching fractions $Br(
\bar B^0\to D^{*}_{s0}(2317)^+K^-)\times Br(D^{*}_{s0}(2317)^+\to D^+_s\pi^0)$
\footnote{Recently, the absolute branching fraction of $D^{*}_{s0}(2317)^+\to D^+_s\pi^0$ has been measured
by the BESIII Collaboration as $1.00^{+0.00}_{-0.14}\pm0.14$ \cite{bes3}.}, which
is given as $(5.3^{+1.5}_{-1.3}\pm0.7\pm1.4)\times10^{-5}$\cite{belle4}. After rescaling
the branching ratio of the decay $D^+\to\phi\pi^+$, PDG reported
$Br(
\bar B^0\to D^{*}_{s0}(2317)^+K^-)\times Br(D^{*}_{s0}(2317)^+\to D^+_s\pi^0)=(4.2^{+1.4}_{-1.3}\pm0.4)\times10^{-5}$ \cite{pdg16}.
Then the Belle Collaboration improved the measurement for the decay $\bar B^0\to D^{*}_{s0}(2317)^+K^-$ and renewed the branching
ratio as $(3.3\pm0.6\pm0.7)\times10^{-5}$\cite{belle5}, where the authors concluded that the branching ratio for
this pure annihilation decay is of the same order of magnitude as $Br(\bar B^0\to D_s^+K^-)$, which is measured as
$(2.7\pm0.5)\times10^{-5}$ \cite{pdg16}. Although the decay $\bar B^0\to D^{*}_{s0}(2317)^+K^-$ has not been measured accurately by experiment, we believe that our prediction
is reasonable.

It is helpful to define the following ratios based on the factorization assumption:
\be
R_1&=&\frac{Br(\bar B^0_s\to D^{*+}_{s0}\pi^-)}{Br(\bar B^0_s\to D^{*+}_{s0}K^-)}\approx\left|\frac{V_{ud}f_\pi}{V_{us}f_K}\right|^2\approx12.4,\\
R_2&=&\frac{Br(\bar B^0_s\to D^{*+}_{s0}\rho^-)}{Br(\bar B^0_s\to D^{*+}_{s0}K^{*-})}\approx\left|\frac{V_{ud}f_\rho}{V_{us}f_{K^*}}\right|^2\approx17.4,
\en
which are consistent with the results given by our predictions.

In the following, we list the branching ratios for the CKM suppressed decays $B_{(s)}\to D^{*}_{s0}(2317)P$ as following
\be
Br(\bar B^0_s\to D^{*}_{s0}(2317)^-K^+)&=&(6.86^{+2.60+0.29+0.45}_{-1.79-0.20-0.43})\times10^{-6},\label{dp1}\\
Br(\bar B^0\to D^{*}_{s0}(2317)^-\pi^+)&=&(6.91^{+2.93+0.45+0.43}_{-1.95-0.44-0.30})\times10^{-6},\\
Br(B^-\to D^{*}_{s0}(2317)^-\pi^0)&=&(3.72^{+1.59+0.23+0.25}_{-1.04-0.16-0.23})\times10^{-6},\\
Br(B^-\to D^{*}_{s0}(2317)^-\eta)&=&(6.30^{+2.17+0.41+0.44}_{-1.52-0.40-0.29})\times10^{-7},\\
Br(B^-\to D^{*}_{s0}(2317)^-\eta')&=&(4.17^{+1.50+0.33+0.27}_{-1.05-0.18-0.27})\times10^{-7},\\
Br(\bar B^0\to D^{*}_{s0}(2317)^-K^+)&=&(5.99^{+0.56+0.33+0.39}_{-0.60-0.31-0.38})\times10^{-9},\\
Br(B^-\to D^{*}_{s0}(2317)^-K^0)&=&(0.82^{+0.17+0.15+0.06}_{-0.15-0.09-0.05})\times10^{-9}\label{dp2},
\en
where the first uncertainty comes from the $w_{b}=0.4\pm0.04(0.5\pm0.05)$ for $B(B_s)$ meson, the second error is
from the hard scale-dependent uncertainty, which we vary from $0.75t$ to $1.25t$,
and the third one is from the CKM matrix elements.
For these CKM suppressed decays, the
factorizable emission diagrams (where $D^{*}_{s0}(2317)^-$ meson is emitted from the weak vertex) are the color favored ones with the
Wilson coefficients $a_1=C_2+C_1/3$, while the nonfactorizable
emission diagrams are highly suppressed by the Wilson coefficient
$C_1/3$. This means that the dominant amplitudes are nearly
proportional to the product of $D^*_{s0}(2317)$ meson decay constant
and a $B$ to light meson form factor. Unfortunately, the decay
constant of the scalar meson for vector current is small, which is
defined as $\langle0|\bar s\gamma_\mu
c|D^*_{s0}(P)\rangle=f_{D^*_{s0}}P_\mu$. This vector current decay
constant $f_{D^*_{s0}}$ can be related with the scale-dependent
scalar one $\tilde f_{D^*_{s0}}$ by equation of motion \be
f_{D^*_{s0}}=\frac{(m_c-m_s)}{m_{D^*_{s0}}}\tilde f_{D^*_{s0}}, \en
where $m_{c(s)}$ is the current quark $c(s)$ mass and $\tilde
f_{D^*_{s0}}$ defined as $\langle|\bar
sc|D^*_{s0}(P)\rangle=m_{D^*_{s0}}\tilde f_{D^*_{s0}}$.
$\tilde f_{D^*_{s0}}=(225\pm25)$ MeV has been determined from the two-point QCD sum rules.
If taking $m_c=1.275$GeV, $m_s=0.096$GeV, $m_{D^*_{s0}}=2.3177$ GeV \cite{pdg16}, one can find that
$f_{D^*_{s0}}=0.11$ GeV. So we can speculate that the branching ratio of the decay  $\bar B^0\to D^{-}_{s}\pi^+$
should be much larger than that of $\bar B^0\to D^{*}_{s0}(2317)^-\pi^+$. This is indeed the case: If we replace
the decay constant, the mass and the wave functions of the scalar meson $D^{*-}_{s0}$ with
those of the pseudoscalar meson $D^-_s$ in the calculation program, we find that the branching ratio
$Br(\bar B^0\to D^{-}_{s}\pi^+)=(27.6^{+8.23}_{-7.65})\times10^{-5}$, which is consistent with the current experimental
value $(21.6\pm2.6)\times10^{-5}$ \cite{pdg16} within errors. Since the form factors of $B\to V$ are a litter large,
one can expect that these tree operator dominant decays $B\to D^{*-}_{s0}V$ have a larger branching
ratios than those of $B\to  D^{*-}_{s0}P$ decays. While this conclusion is not satisfied to
the pure annihilation type decays.

For the decays $\bar B^0\to D^{*-}_{s0}\pi^+, B^-\to D^{*-}_{s0}\pi^0$, their branching ratios are sensitive
to the form factor $B\to \pi$. If using the Gegenbauer coefficients $a^\pi_2=0.44, a^\pi_4=0.25$,
we will get a reasonable form factor $F^{B\to \pi}(0)=0.22$, which is larger than $F^{B\to \pi}(0)=0.18$ obtained
by using the updated Gegenbauer coefficients $a^\pi_2=0.115, a^\pi_4=-0.015$. Corresponding to the smaller form factor,
the branching ratios of the
decays $\bar B^0\to D^{*-}_{s0}\pi^+, B^-\to D^{*-}_{s0}\pi^0$ will have a noticeable decrement and become
$Br(\bar B^0\to D^{*-}_{s0}\pi^+)=3.78\times10^{-6}, Br(B^-\to D^{*-}_{s0}\pi^0)=2.05\times10^{-6}$. It is
similar for the decays $B^-\to D^{*-}_{s0}\eta^{(\prime)}$.
While for the CKM favored decay $\bar B^0_s\to D^{*+}_{s0}\pi^-$, the branching ratio is not sensitive to the Gegenbauer coefficients for $\pi$ meson
wave functions. The difference of the branch ratios by using
these two group Gegenbauer coefficients is only about $4\%$ .

Similarly, the branching ratios of the decays $B_{(s)}\to D^{*}_{s0}(2317)^-V$
are calculated as:
\be
Br(\bar B^0_s\to D^{*}_{s0}(2317)^-K^{*+})&=&(7.97^{+2.56+0.49+0.52}_{-2.14-0.64-0.51})\times10^{-6},\\
Br(\bar B^0\to D^{*}_{s0}(2317)^-\rho^{+})&=&(1.61^{+0.64+0.16+0.11}_{-0.46-0.10-0.10})\times10^{-5},\\
Br(B^-\to D^{*}_{s0}(2317)^-\rho^0)&=&(8.70^{+3.42+0.51+1.53}_{-2.34-0.56-0.52})\times10^{-6},\\
Br(B^-\to D^{*}_{s0}(2317)^-\omega)&=&(5.44^{+2.16+0.52+0.36}_{-1.48-0.36-0.30})\times10^{-6},\\
Br(B^-\to D^{*}_{s0}(2317)^-\phi)&=&(1.74^{+0.67+0.33+0.11}_{-0.50-0.26-0.11})\times10^{-8},\\
Br(\bar B^0\to D^{*}_{s0}(2317)^-K^{*+})&=&(6.38^{+1.25+0.18+0.41}_{-1.02-0.48-0.41})\times10^{-9},\\
Br(B^-\to D^{*}_{s0}(2317)^-K^{*0})&=&(0.73^{+0.19+0.10+0.04}_{-0.21-0.08-0.05})\times10^{-9},
\en
where the errors are the same as ones given in Eqs.(\ref{dp1})-(\ref{dp2}).
\begin{table}
\caption{The amplitudes from the nonfactorizable and factorizable annihilation Feynman diagrams, which denote as NFAA and FAA, respectively.
For each amplitude, the value has been given, together with the corresponding Wilson coefficient (WC). }
\begin{center}
\begin{tabular}{c|c|c|c|c|c}
\hline\hline \multirow{2}{*}{Modes} & \multicolumn{2}{|c|}{NFAA}&
\multicolumn{2}{|c|}{FAA} &\multirow{2}{*}{Total}\\ \cline{2-5}
               &WC&value&WC&value& \\ \cline{1-6}

$B^-\to D^{*-}_{s0}K^0(\times10^{5})$  &\multirow{2}{*}{$C_1$}&$-3.99-i2.75$&\multirow{2}{*}{$C_2+C_1/3$}&$1.72-i0.57$&$-2.13-i3.31$\\
$B^-\to D^{*-}_{s0}K^{*0}(\times10^{5})$  &&$-1.02-i4.60$&&$1.38+i0.87$&$-0.50-i3.72$\\
\hline
$\bar B^0\to D^{*-}_{s0}K^+(\times10^{5})$ &\multirow{2}{*}{$C_2$}&$8.50-i8.22$&\multirow{2}{*}{$C_1+C_2/3$}&$-0.50-i0.35$&$8.00-i8.57$\\
$\bar B^0\to D^{*-}_{s0}K^{*+}(\times10^{5})$ &&$-8.80-i13.54$&&$0.07-i0.11$&$-0.26-i11.5$\\
\hline
$\bar B^0\to D^{*+}_{s0}K^-(\times10^{3})$&\multirow{2}{*}{$C_2$}&$4.46-i4.44$&\multirow{2}{*}{$C_1+C_2/3$}&$-0.32-i0.16$&$4.14-i4.60$\\
$\bar B^0\to D^{*+}_{s0}K^{*-}(\times10^{3})$&&$4.83-i5.47$&&$-0.18-i0.06$&$4.65-i5.53$\\
\hline\hline
\end{tabular}\label{tab2}
\end{center}
\end{table}

The pure annihilation decays have the smallest branching
ratios both for the CKM allowed and the CKM suppressed ones. In Table \ref{tab2}, we list the contributions from
the nonfactorizable annihilation amplitudes (NFAA) and the factorizable annihilation amplitudes (FAA), where
the Wilson coefficients have been included. One can find that the nonfactorizable contributions are more important
than the factorizable ones. Even the FAA with the large Wilson Coefficient ($a_1=C_2+C_1/3$) also has smaller value
because of the destructive interference between the pair of factorizable annihilation Feynman diagrams in each channel,
such as Fig.1(g) and Fig.1(h).
For example, in the decay $B^-\to D^{*-}_{s0}K^0$ both of the two factorization annihilation amplitudes have large imaginary parts in magnitude but with opposite
signs: One is $3.71\times10^{-5}$, the other is $-4.28\times10^{-5}$, so the imaginary part of the total FAA
becomes $-5.7\times10^{-6}$ given in Table \ref{tab2}.
\section{Conclusion}\label{summary}
In summary, we investigate the branching ratios of the decays $B_{(s)}\to D^*_{s0}(2317)P(V)$ within pQCD approach by
assuming $D^*_{s0}(2317)$ as a $\bar cs$ scalar meson. For the CKM favored decays, their branching fractions are larger
than $10^{-5}$, even for the pure annihilation type channels. Our predictions are consistent well with the results given by
the light cone sum rules approach. So we consider that these decays can be measured at the running LHCb and the forthcoming SuperKEKB.
We may shed light on the nature of the meson $D^*_{s0}(2317)$ by combining with the future data and the theoretical predictions: If they
are consistent with each other, one can conclude that this charmed-strange meson is composed (mainly) of $\bar cs$.
Otherwise, some other component or the $DK$ threshold effect in meson-mseon scattering may be important to the dynamic mechanism
for the $D^*_{s0}(2317)$ production. As for the CKM suppressed decays, their branching ratios are usually at $10^{-6}$ order. While
the branching fraction for the decay $\bar B^0\to D^{*}_{s0}(2317)^-\rho^+$ reaches up to $1.61\times10^{-5}$. As to
the pure annihilation type decays $B^-\to D^{*}_{s0}(2317)^-\phi, \bar B^0\to D^{*}_{s0}(2317)^-K^+(K^{*+})$ and
$B^-\to D^{*}_{s0}(2317)^-K^0(K^{*0})$, their branching fractions drop to as low as $10^{-10}\sim 10^{-8}$. Here the decay
$B^-\to D^{*}_{s0}(2317)^-\phi$ has the larger branching ratio because of the large CKM matrix element $V_{cs}$. The branching ratios of
the decays $\bar B^0\to D^{*}_{s0}(2317)^-K^+(K^{*+})$ are larger than those of $B^-\to D^{*}_{s0}(2317)^-K^0(K^{*0})$ because of
owning the larger nonfactorizable annihilation amplitudes. For these pure annihilation type decays, the magnitudes of
the nonfactorizable amplitudes are generally larger than those of factorization amplitudes. It is because there exists the
destructive interference between the pair of factorization amplitudes in each decay mode. If this type of pure annihilation
decay is
observed by the future experiments with larger branching fractions than our predictions, it may indicate that
some new physics contributes to these decays.
\section*{Acknowledgment}
This work is supported by the National Natural Science
Foundation of China under Grant No. 11347030, 11847097, and by the Program of
Science and Technology Innovation Talents in Universities of Henan
Province 14HASTIT037. One of us (N. Wang) is supported by the Science Research Fund Project for the High-Level Talents of Henan University of Technology 0004/31401151.
\appendix
\section{}
\be
t_a&=&\max(\sqrt{x_2}m_B,1/b_1,1/b_2),\\
t_b&=&\max(\sqrt{x_1}m_B,1/b_1,1/b_2),\\
t^\prime_{a}&=&\max(\sqrt{x_3(1-r^2_{D^*_{s0}})}m_B,1/b_1,1/b_3),\\
t^\prime_b&=&\max(\sqrt{x_1(1-r^2_{D^*_{s0}})}m_B,1/b_1,1/b_3),\\
t_{c,d}&=&\max(\sqrt{x_1x_2}m_B,\sqrt{|A^{2}_{c,d}|}m_B,1/b_1,1/b_2),\\
t^{\prime}_{c,d}&=&\max(\sqrt{x_1x_3(1-r_{D^*_{s0}}^2)}m_B,\sqrt{|A^{\prime2}_{c,d}|}m_B,1/b_1,1/b_3),\\
t_{e,f}&=&\max(\sqrt{x_2(1-x_3)(1-r_{D^*_{s0}}^2)}m_B,\sqrt{|L^2_{e,f}|},m_B,1/b_1,1/b_3),\\
t^{\prime}_{e,f}&=&\max(\sqrt{x_2x_3(1-r_{D^*_{s0}}^2)}m_B,\sqrt{|L^{\prime2}_{e,f}|},m_B,1/b_1,1/b_3),\\
t_{g}&=&\max(\sqrt{(1-x_3)(1-r^2_{D^*_{s0}})}m_B,1/b_2,1/b_3), \\
t_{h}&=&t^{\prime}_{g}=\max(\sqrt{x_2(1-r^2_{D^*_{s0}})}m_B,1/b_2,1/b_3),\\
t^{\prime}_{h}&=&\max(\sqrt{x_3(1-r^2_{D^*_{s0}})}m_B,1/b_2,1/b_3),
\en
where the definitions of $A^{(\prime)}_{c,d},L^{(\prime)}_{e,f}$ are listed in
Eqs.(\ref{ac1}),(\ref{ad1}),(\ref{lef1}),(\ref{lef2}),(\ref{ac2})-(\ref{lefp2}).
And the $S_j(t) (j=B,D_{D^*_{s0}},P)$ functions in Sudakov form factors in Eq.(\ref{suda1}), Eq.(\ref{suda2}),
Eq.(\ref{suda3}) and Eq.(\ref{suda4}) are given as
\be
S_B(t)&=&s(x_1\frac{m_B}{\sqrt2},b_1)+2\int^t_{1/b_1}\frac{d\bar\mu}{\bar\mu}\gamma_q(\alpha_s(\bar\mu)),\\
S_{D^*_{s0}}(t)&=&s(x_3\frac{m_B}{\sqrt2},b_3)+2\int^t_{1/b_3}\frac{d\bar\mu}{\bar\mu}\gamma_q(\alpha_s(\bar\mu)),\\
S_{P}(t)&=&s(x_2\frac{m_B}{\sqrt2},b_2)+s((1-x_2)\frac{m_B}{\sqrt2},b_2)+2\int^t_{1/b_2}\frac{d\bar\mu}{\bar\mu}\gamma_q(\alpha_s(\bar\mu)),
\en
where the quark anomalous dimension $\gamma_q=-\alpha_s/\pi$, and the expression of the $s(Q,b)$ in one-loop running coupling
coupling constant is used
\be
s(Q,b)&=&\frac{A^{(1)}}{2\beta_1}\hat{q}\ln(\frac{\hat{q}}{\hat{b}})-\frac{A^{(1)}}{2\beta_1}(\hat{q}-\hat{b})
+\frac{A^{(2)}}{4\beta^2_1}(\frac{\hat{q}}{\hat{b}}-1)\non &&-\left[\frac{A^{(2)}}{4\beta^2_1}-\frac{A^{(1)}}{4\beta_1}\ln(\frac{e^{2\gamma_E-1}}{2})\right]
\ln(\frac{\hat{q}}{\hat{b}}),
\en
with the variables are defined by $\hat{q}=\ln[Q/(\sqrt2\Lambda)], \hat{q}=\ln[1/(b\Lambda)]$ and the coefficients $A^{(1,2)}$
and $\beta_{1}$ are
\be
\beta_1&=&\frac{33-2n_f}{12},A^{(1)}=\frac{4}{3},\\
A^{(2)}&=&\frac{67}{9}-\frac{\pi^2}{3}
-\frac{10}{27}n_f+\frac{8}{3}\beta_1\ln(\frac{1}{2}e^{\gamma_E}),
\en
here $n_f$ is the number of the quark flavors and $\gamma_E$ the Euler constant.


\begin{thebibliography}{99}
\bibitem{babar1}
B. Aubert {\it et al.}[BABAR Collaboration], \prl {\bf90},242001 (2003).
\bibitem{babar2}
B. Aubert {\it et al.}[BABAR Collaboration], \prl {\bf93},181801 (2004).
\bibitem{cleo}
D. Besson {\it et al.}[CLEO Collaboration], \prd {\bf68}, 032002 (2003).
\bibitem{belle}
P. Krokovny {\it et al.}[Belle Collaboration], \prl {\bf91}, 262002 (2003).
\bibitem{god}
S. Godfrey and N. Isgur, \prd {\bf32}, 189 (1985); S. Godfrey and R. Kokoski, \prd {43}, 1679 (1991);
J. Zeng, J. W. Van Orden and W. Roberts, \prd {\bf52}, 5229 (1995); D. Ebert, V.O. Galkin and R.N. Faustov
,\prd{\bf57}, 5663 (1998).
\bibitem{kala}
Y.S. Kalashnikova, A.V. Nefediev and Y.A. Simonov, \prd {\bf64}, 014037 (2001); M.Di Pierro and E.Eichten, \prd {\bf64},
114004 (2001).
\bibitem{bali}
G.S. Bali, \prd {\bf68}, 071501 (2003); A. Dougall {\it et al.} [UKQCD Collaboration], \plb {\bf569}, 41 (2003).
\bibitem{haya}
A. Hayashigaki, K. Terasaki, arXiv:hep-ex/0411285.
\bibitem{narison}
S. Narison, \plb {\bf605}, 319, 2005.
\bibitem{bes3}
M. Ablikim, {\it et al.}[BESIII Collaboration], \prd {\bf97}, 051103 (2018).
\bibitem{barn}
T. Barnes, F. E. Close, and H.J. Lipkin, \prd {\bf68}, 054006 (2003).
\bibitem{chenyq}
Y.Q. Chen, and X.Q. Li, \prl{\bf93}, 232001 (2004).
\bibitem{guo}
F.K. Guo, P.N. Shen, H.C. Chiang, R.G. Ping, and B.S. Zou, \plb{\bf641}, 278 (2006).
\bibitem{faes}
A. Faessler, T. Gutsche, V.E. Lyubovitskij, and Y.L. Ma, \prd {\bf76},014005 (2007).
\bibitem{guo1}
F.K.Guo, C. Hanhart, and U.G. Meissner, Eur.Phys.J.A {\bf40}, 171 (2009).
\bibitem{hycheng}
H.Y. Cheng and W.S. Hou, \plb{\bf566}, 193 (2003).
\bibitem{tera}
K. Terasaki, \prd{\bf68}, 011501(R) (2003).
\bibitem{dmi}
V. Dmitrasinovic, \prl {\bf94}, 162002 (2005).
\bibitem{jrzhang}
J.R. Zhang, \plb{\bf789}, 432 (2019).
\bibitem{beve}
E.V. Beveran and G. Rupp, \prl {\bf91}, 012003 (2003).
\bibitem{mohler}
D. Mohler, C.B. Lang, L.Leskovec, S.Prelovsek, and R.M. Woloshyn, \prl{\bf111},222001 (2013).
\bibitem{liul}
L.Liu,K. Orginos, F.K. Guo, C. Hanhart, and U.G. Meissner, \prd{\bf87}, 014508 (2013).
\bibitem{zqzhang1}
Z.Q. Zhang, S.Y.Wang, X.K. Ma, \prd{\bf93}, 054034 (2016).
\bibitem{zqzhang2}
Z.Q. Zhang, S.J. Wang, L.Y.Zhang, Chin. Phys. C {\bf37}, 043103 (2013).
\bibitem{zqzhang3}
H.Y. Cheng, C. K. Chua, K.C. Yang, Z.Q. Zhang, \prd{\bf87}, 114001 (2013).
\bibitem{zqzhang4}
Z.Q.Zhang, Eur. Phys. Lett. {\bf97}, 11001 (2012).
\bibitem{zou}
Z.T. Zou, Y. Li, X. Liu, Eur. Phys. J. C{\bf77}, 870 (2017).
\bibitem{lirh}
R.H. Li, C.D.Lu, and Y.M. Wang, \prd{\bf80}, 014005 (2009).
\bibitem{faus}
R.N. Faustov, V.O. Galkin, \prd{\bf87},034033 (2013).
\bibitem{albert}
C. Albertus, \prd {\bf89}, 065042 (2014).
\bibitem{cdlu}
C.D. Lu, M.Z. Yang, Eur. Phys. J. C{\bf28}, 515 (2003).
\bibitem{ali}
A.Ali, {\it et al.}, \prd{\bf76}, 074018 (2007).
\bibitem{chench}
C.H. Chen, \prd{\bf68}, 114008 (2003).
\bibitem{wfun1}
V.L. Chernyak and A.R. Zhitnitsky, Phys. Rept.{\bf112}, 173 (1984).
\bibitem{wfun2}
A.R. Zhitnitsky, I.R. Zhitnitsky and V.L. Chernyak, Sov.J. Nucl. Phys. {\bf41}, 284 (1985).
\bibitem{wfun3}
V.M.Braun and I.E. Filyanov, Z.Physik C{\bf44}, 157 (1989).
\bibitem{pball}
P.Ball, JHEP {\bf9809},005 (1998); JHEP {\bf9901}, 010 (1999).
\bibitem{pball1}
P.Ball and R. Zwicky, \prd{\bf71}, 014029 (2005); JHEP {\bf0604}, 046 (2006).
\bibitem{pball2}
P. Ball and G.W.Jones, JHEP {\bf0703}, 069 (2007).
\bibitem{hnli}
H.-n. Li, Phys.Lett.B {\bf622}, 63 (2005).
\bibitem{feldmann}
T.Feldmann, P. Kroll and B. Stech, \prd{\bf58}, 114006 (1998); \plb{\bf449}, 339 (1999).
\bibitem{pdg16}
C. Patrignani {\it et al.}[Particle Data Group Collaboration], Chin.Phys.C {\bf40}, 100001 (2016).
\bibitem{belle4}
A. Drutskoy, {\it et al.}[Belle Collaboration], \prl {\bf94}, 061802 (2005).
\bibitem{belle5}
K. Abe, {\it et al.}[Belle Collaboration], arXiv:hep-ex/0507064.
\end{thebibliography}
\end{document}